\documentclass[aps,twocolumn,showpacs,pra]{revtex4-1}
\usepackage{graphicx,epstopdf}
\usepackage{amsmath}
\usepackage{amssymb}
\usepackage{color}
\usepackage{float}
\usepackage{epsfig}
\usepackage{epstopdf}
\usepackage{palatino}
\usepackage{times}
\usepackage{enumitem}
\usepackage{tcolorbox}
\usepackage{bmpsize}
\usepackage[normalem]{ulem}

\usepackage[colorlinks=true, citecolor=blue, urlcolor=blue ]{hyperref}
\input{epsf}
\begin{document}
\title {Trends of information backflow in disordered spin chains}
\author{Sudipto Singha Roy\(^{1}\), Utkarsh Mishra\(^{2}\), and Debraj Rakshit\(^{3}\)}
\affiliation{\(^1\) Instituto de F{\'i}sica T{\'e}orica UAM/CSIC,  C/ Nicol{\'a}s Cabrera 13-15, Cantoblanco, 28049  Madrid, Spain\\
\(^2\) Institute of Fundamental and Frontier Sciences, University of Electronic Science and Technology of China, Chengdu 610051, China\\
\(^3\)ICFO--Institut de Ci{\'e}ncies Fot{\'o}niques, The Barcelona Institute of Science and Technology, 08860 Castelldefels, Spain}
\begin{abstract}
We investigate the trends of information backflow  associated with the dynamics of a sub-part of a disordered spin-1/2 transverse field  Heisenberg chain for different  regimes of the Hamiltonian. Towards this aim, the decay profile of bipartite entanglement shared between a probe-qubit and a system-qubit (sub-part) of the chain is monitored in time. A clear shift in the trends of the decay profiles of the bipartite entanglement from monotonic in the low-disorder limit to non-monotonic in the moderately large disorder limit occurs due to strong information backflow from the environment (complementary-part) to the system-qubit. A connection between environmental interruption caused by the information backflow and the disorder strength is established by examining the entanglement revival frequencies. The growth patterns of the revival frequencies in the localized phase plays an instrumental role to effectively distinguish an interacting system (many-body localized) from its non-interacting (Anderson localized) counterpart.
\end{abstract}
\maketitle

\section{Introduction}

Investigations of  quantum properties associated with disordered  systems have gained much attention in last few decades~\cite{disorder_bunch1,disorder_bunch2,disorder_bunch3,disorder_group1,
disorder_group2,disorder_group3,mbl1,mbl2,mbl3,mbl4}. This is primarily due to the fact that disorder is unavoidable in real materials and nowadays can also be engineered  in a well-controlled setting in  laboratories~\cite{disorder_exp1,disorder_exp2,disorder_exp3,disorder_exp4}. Moreover, presence of disorder gives rise to many interesting phenomena, which are in sharp contrast to the behavior observed in their homogeneous counterparts. The list includes order-from-disorder~\cite{disorder_group1,disorder_group2,disorder_group3},  enhancement of quantum correlation length~\cite{disorder_group1}, strong violation of area-law~\cite{disorder_bunch3}, and    disorder-induced localization~\cite{disorder_bunch1, mbl1,mbl2,mbl3,mbl4,disorder_quench1,disorder_quench2,disorder_quench3,markovian_qubit1,
markovian_qubit2,thermalization1,thermalization2,thermalization3}. 
The study of
entanglement entropy dynamics  in disordered quantum systems has offered new understandings on many-body localization~\cite{disorder_quench1,disorder_quench2,disorder_quench3,markovian_qubit1,markovian_qubit2}.   In this regard, study the dynamics of quantum properties in the disordered quantum systems exploiting the tools of open quantum systems provide a different albeit interesting route to understand the microscopic properties of the systems in more detail \cite{ent_dis_dynmcs}. This forms a central theme of investigation of the present work.\

 In the realm of the theory of open quantum systems,   the coupling between system and environment in Markovian dynamics can be assumed such that the environment is
memoryless and uncorrelated with the system. This is a manifestation of the fact that the environment-induced changes to the system dynamics are slow relative to the typical correlation time of the environment, or rather rephrasing, there is a sharp detachment between the typical correlation timescale of the fluctuations and the timescale of the evolution of the system~\cite{Born_Markov}. As a result, there is always loss of information from the system to the environment, but not the other way around.  However, in reality, the  actual dynamics of an  open quantum system often deviates significantly than this  idealized scenario and it is of practical interest (e.g. experimental implementations) to consider non-Markovian evolution, where there are instances at which  the memory effect causes revivals of the quantum properties of the system, which is commonly known as  backflow of information from the environment to the system~\cite{ref:book}. The memory effects associated with the non-Markovian dynamics can be attributed as a reminder to the system about its past. In other words, during the course of evolution, the environment stores the information about the initial conditions of the system for a while and later information flows back from the environment, which reminds the system about its past. Despite the fact that the concept of non-Markovianity is well established in the classical case~\cite{classical_Marko_non_Marko}, its quantum generalization still remains subtle and ambiguous. In recent years, significant attempts have been made  for providing computable measures of non-Markovianity in a precise manner. The most customary one is the  characterization of   non-Markovianity based  on  the   non-monotonic time evolution of some quantum information measures, viz. quantum entanglement~\cite{non_Marko_ent}, quantum mutual information\cite{non_Marko_MI}, quantum Fisher information~\cite{non_Marko_FI},  quantum interferometric power~\cite{non_Marko_Interfero}, quantum coherence~\cite{non_Marko_coherence} etc.
 \\

\begin{figure}[h!]
\includegraphics[width=3.4 in,angle=00]{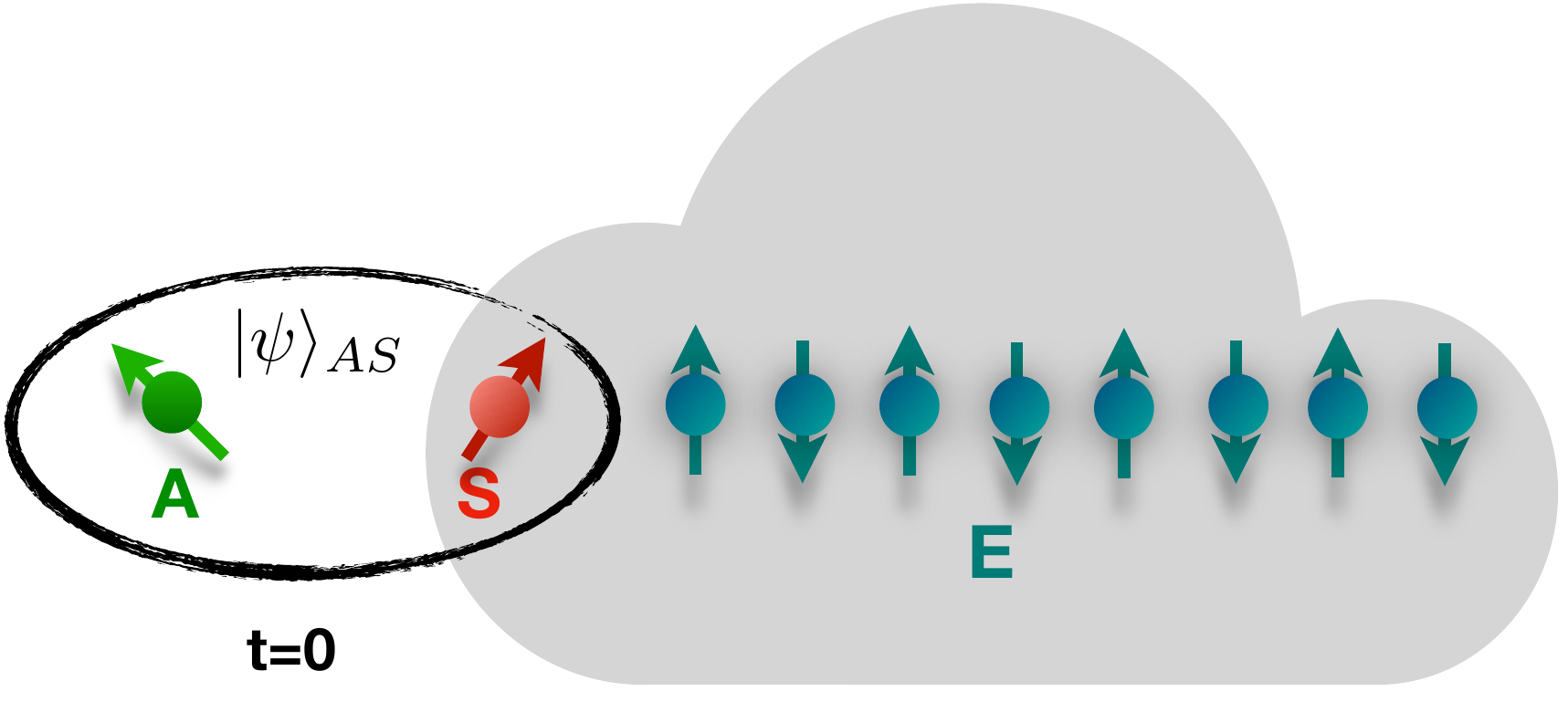}
\caption{Schematic diagram of the system-environment ($S: E$) bipartition of an $N$-qubit chain. One of the sites of the spin-chain is considered as a system ($S$) while the rest of the chain serves as an environment ($E$). Initially ($t=0$), the system-qubit remains in a maximally entangled state, $|\psi\rangle_{AS}$, with probe-qubit $A$.}
\label{schematic}
\end{figure}

In this work, we propose a framework to study the memory effects associated with the dynamics of the subsystems of a many-body quantum system. Towards this aim, we divide the whole many-body quantum state into two parts, system-part and environment-part and study the effects of environmental influences on the dynamics of the system-qubits. As a prototypical model, we consider the Heisenberg chain with the disordered transverse field~\cite{mbl2,mbl3,mbl4, ETH_MBL_others2,
ETH_MBL_others3,ETH_MBL_others7,MBL_ETH_entanglement1,
MBL_ETH_entanglement2,MBL_ETH_entanglement3,Alet1}. Dividing the total $N$-qubit system in single-qubit  `system', ${S}$, and $(N-1)$-qubit `environment', ${E}$, we characterize the dynamical map acting on the single-qubit system-part   from the decay profile of initial bipartite entanglement shared between the system-qubit and a probe-qubit, $A$ (see Fig.~\ref{schematic}).
We note that the decay profile of the initial bipartite entanglement between $A$ and $S$ has a strong dependence on the disorder strength. For instance, at  small values of the disorder strength, when $SE$ remains in the ergodic phase \cite{mbl1,mbl2,mbl3,mbl4}, the initial entanglement decays to zero monotonously,  implying a  information flow from $S$ to $E$, which causes complete erasure of its initial memory. Subsequently, at a higher value of the disorder strength, corresponding to the localized phase of $SE$, strong non-Markovian nature of the dynamical map becomes evident, which is characterized via the non-monotonic decay of  bipartite entanglement.   In order to obtain a quantitative understanding of the correspondence between the environmental interruptions and the disorder strength, we propose a measure by counting the frequency of the \emph{near-perfect} revivals of the  bipartite entanglement. Interestingly, this newly introduced measure provides a clear distinction between  two different types of localized phases those appear in absence  (Anderson localization) and in presence (many-body localization)  of interaction, even for a very small value of the interaction strength.\

In this regard, we would like to mention here that there has been considerable interest in studying open system dynamics in the presence of disordered environments \cite{nonmarkovianity_Anderson1,nonmarkovianity_Anderson2,
nonmarkovianity_Anderson3,nonmarkovianity_Anderson4}. In particular, in \cite{nonmarkovianity_Anderson1,nonmarkovianity_Anderson2}
it has been shown that the dynamics of a two-label system coupled to an array of cavities with the static disorder, acting as the environment, exhibit information backflow. In all these cases, the localization property that has been addressed is Anderson localization in nature. However, in our case, as  the model we consider is an interacting one, this additionally provides a scope to find a more fine-grained characterization of localized phases, in terms of  information backflow.

In the following sections, after briefly introducing the model, we discuss the methodology undertaken and elaborate on our main findings.\

\section{ Model and Methodology}

\label{model}

In general, for a real dissipative system, where the system usually interacts with an infinitely large environment,   the degrees of freedom of the environment-part remain inaccessible. Hence, it is not always an easy task to characterize its influences on the dynamics of the system-qubits. However, in case of a system and environment, both consisting of a finite number of sites,  the characterization of system dynamics becomes straightforward,  yet the perspective of understanding the system exploiting the theory of open quantum systems opens up. This brings the possibility of providing new insightful properties related to the system dynamics.
In this work,  we consider the spin 1/2 Heisenberg model in one-dimensional with a random field along the $z$-direction. The Hamiltonian is given by
\begin{eqnarray}
H=\sum_{i=1}^{N-1} \Big[J(S^x_iS^x_{i+1}+S^y_iS^y_{i+1})+\Delta S^z_i S^z_{i+1}\Big
]+ \sum_{i=1}^{N}h_iS^z_i,\nonumber\\
\label{Ham_main}
\end{eqnarray}
where $h_i$ are independent random variables at each site $i$, each with a probability distribution that is uniform in $[-h,h]$,  $J$ is the coupling constant along the $x$- and $y$- directions and $\Delta$ is the same for $z$- direction.  Moreover, $S^i=\sigma^i/2$, with $i\in \{x,y,z\}$.

 The total Hamiltonian $H$ can be decomposed as

\begin{eqnarray}
H=H_S+H_E+H_I,
\end{eqnarray}
where  $H_S=h_1S^z_1$ is the single-particle Hamiltonian corresponding to a single spin at one of the edges of the spin chain,  $H_I=J (S^x_1S_2^x+S^y_1S_2^y)+\Delta S^z_1S_2^z$ represent the interaction between the edge spin with the nearest-neighbour spin. $H_E$ has  same mathematical form as the Hamiltonian  expressed in Eq.~(\ref{Ham_main}) but with  $N-1$ number of particles.\

For a open quantum system, the evolutions  are generally represented by dynamical  map, $\Lambda_t$, given by~\cite{ref:book}
\begin{eqnarray}
\rho_S(t)=\Lambda_t(\rho_S(0))=\text{Tr}_E(U_{SE}\rho_{SE}(0) U^{\dagger}_{SE}),
\label{map}
\end{eqnarray}
where $U_{SE}$ is a unitary acting on system-environment state $\rho_{SE}$.
In our case,  to investigate the trends of information backflow in different phases of the $N$-qubit system,  we conduct a two-steps analysis.  Firstly, we  consider the initial system state, $\rho_S(0)$, is evolving under a dynamical map, $\Lambda_t$, obtained  using  Eq.~(\ref{map}) for certain choices of the initial environment state $\rho_E(0)$ and the unitary  $U_{SE}=e^{-iHt}$ acting on system-environment state, where $H$ is expressed in Eq.~(\ref{Ham_main}).  We then characterize the properties of the dynamical map $\Lambda_t$ by studying its effects on the dynamics of the quantum entanglement shared between the system and a probe-qubit, $A$. We describe the procedure in details in the forthcoming section.\

\section{Characterization of the dynamical map $\Lambda_t$}
\label{analysis}
To characterize the properties of the dynamical map, $\Lambda_t$, acting on the system-qubits in detail, we consider the external probe-qubit, $A$, is maximally entangled with $S$. The system-probe state has following mathematical form in the computational basis $|\psi\rangle_{AS}=\frac{1}{\sqrt{2}}(|00\rangle+|11\rangle)$.  In order to create this entanglement, one can consider that the qubits were brought together in past and global operations were performed on them~\cite{EPR_creation}.  In addition to this, we consider initially,  the environment is in the state $\rho_E(0)=|01\dots 01\rangle\langle 01\dots 01|_{N-1}$.  It is known that the ergodic properties of the model can be well captured when the middle of the spectrum of the Hamiltonian is considered and therefore any initial state of sufficiently high energy density should yield similar results. However, from our observation we found that among all other choices of the initial environment states, the N{\'e}el-like states are most suitable for distinguishing the quantum revival patterns at different parametric regime of the considered model. Therefore, the initial system-environment joint state can be expressed as, $\rho_{SE}=\frac{\mathbb{I}_S}{2}\otimes |01\dots 01\rangle \langle 01\dots 01|_E$,
where  $\mathbb{I}_S$ is the identity matrix acting on the system's Hilbert space.
 The evolution of the initially maximally entangled system-probe bipartite state, $\rho_{AS}(0)=|\psi\rangle\langle \psi|_{AS}$,  can be written as
\begin{eqnarray}
\rho_{AS}(t)=( \mathbb{I} \otimes  \Lambda_t ) \rho_{AS}(0).
\label{rho_SA}
\end{eqnarray}
Once the bipartite state corresponding to the system and probe-qubit  is evaluated, we can monitor the decay of initial system-probe entanglement with time.

\begin{figure}[h!]
\begin{center}
\includegraphics[width=2.4 in,angle=00]{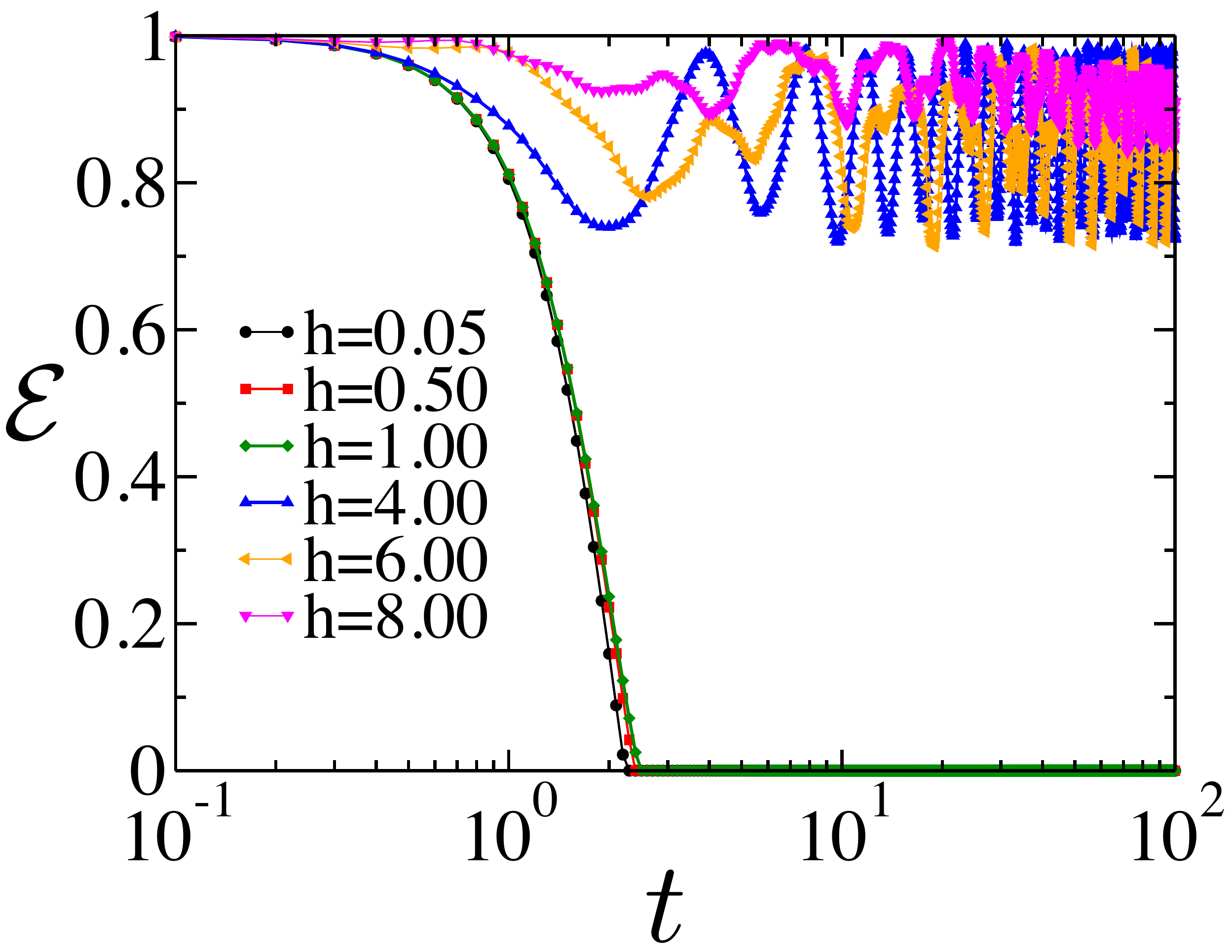}
\caption{(Color online.)  Plot of decay of bipartite entanglement ($\mathcal{E}$) between the system and the probe-qubit with time ($t$), obtained for a single disorder realization, for different values of the disorder strength ($h$). 
The plots are obtained for $J=\Delta=1$ and $N=10$.}
\label{ETH_MBL}
\end{center}
\end{figure}

Figure~\ref{ETH_MBL} depicts the variation of entanglement $\mathcal{E}$ with time $t$, for the time-evolved state given in Eq.~(\ref{rho_SA}), for various strength of the random field and for a single disorder realization with $N=10$.  From the Figure, one can clearly observe that at low values of the disorder strength, the system-probe entanglement decays monotonically from its initial maximum value and eventually goes to zero rapidly. This reveals the fact that the dynamical map acting on the system-qubit is Markovian in nature and we interpret this monotonous decay of entanglement as loss of information from the system to the environment, which causes complete erasure of the initial memory of the system. Therefore, one can argue that when the total $N$-qubit system remains in the ergodic phase, the system-qubit ($S$) becomes more entangled with the bulk-part ($E$) which eventually causes to diminish the quantum correlation between the system and the probe qubit monotonically and  gives rise to a Markovian-like scenario.

Interestingly, the decay profiles of quantum entanglement at moderately large values of the disorder strength differ significantly from the behaviour obtained at low $h$ (see Fig.~\ref{ETH_MBL}). In particular, the decay of entanglement between the system and the probe-qubit exhibits highly non-monotonous behavior, which asserts the fact that during the course of evolution information backflow from the environment to the system occurs in several occasions. In other words, the dynamical map acting on the system-part is non-markovian in nature. The information-flow from the environment to the system essentially keeps reminding the system about its past.   This phenomenon can be thought as a result of certain dynamical decoupling effect that occurs in the localized phase of the model, where the system and the bulk qubit interacts weakly.\\

\begin{figure}[t]
\begin{center}
\includegraphics[width=2.5 in,angle=00]{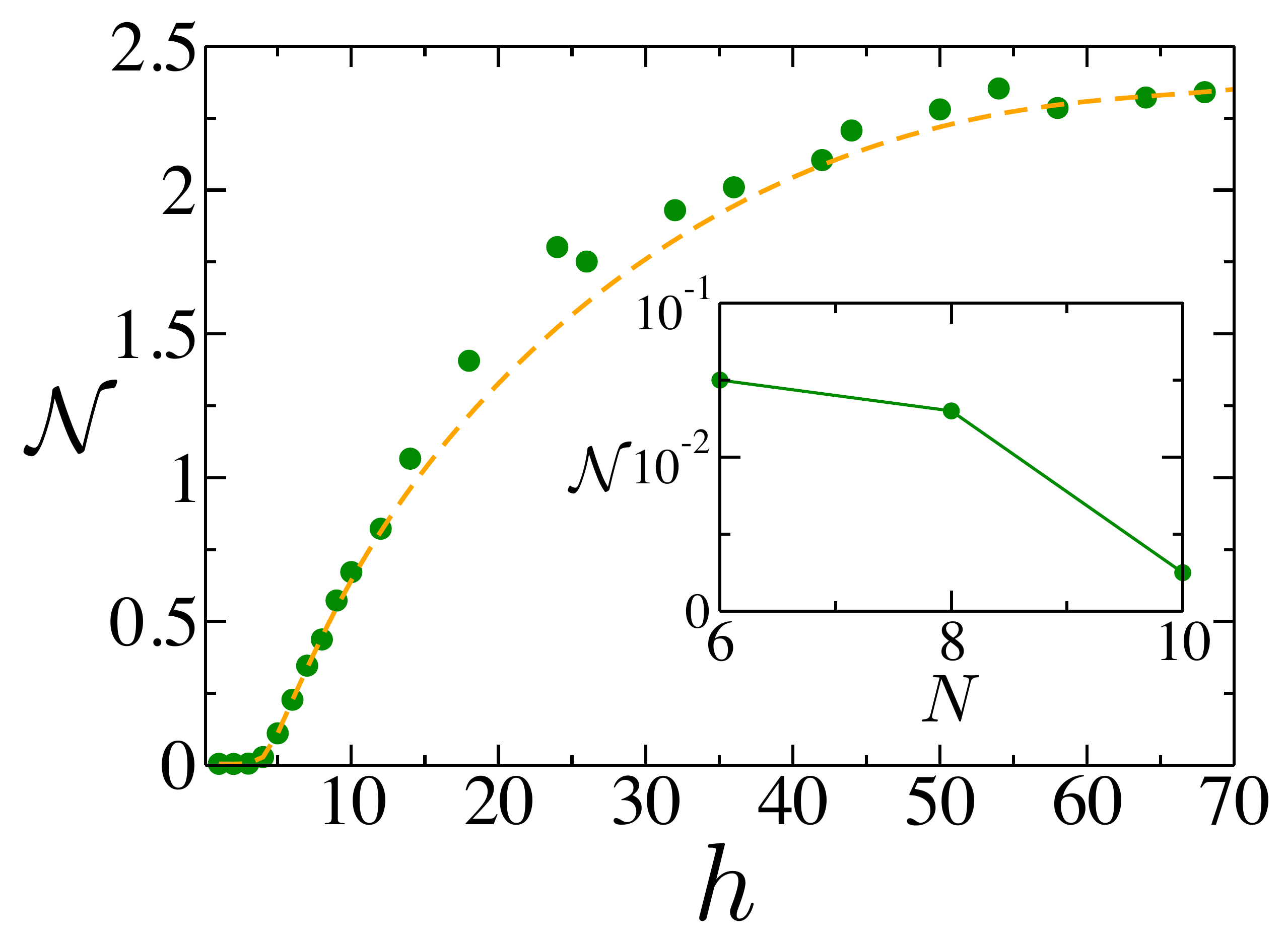}
\caption{(Color online.) Plot of the quenched averaged value of the revival frequency ($\mathcal{N}$) of the bipartite entanglement shared between the system and the probe-qubit with the disorder strength ($h$), obtained for $10^2$ disorder realization, $J=\Delta=1$, and for total time interval $t=100$.  The dashed line serves as a guide to the eye. In the inset, we plot the scaling of $\mathcal{N}$,  for a low value of  disorder strength $h=1$, and for different $N=S+E$ sizes ($N=6, 8, 10$). }
\label{ETH_MBL_Measure}
\end{center}
\end{figure}

In the subsequent part of our analysis, we move one step further and relate  the frequency of environmental interruption  to the disorder strength. Towards this aim,  we count the number of instances during the total period of evolution, at which the bipartite entanglement starts reviving. We argue that the number of revivals accounts for the instances of reminder from the environment to the system about its past.  
On the other hand, a monotonic decay of bipartite entanglement asserts no interruption caused by the environment.
The ``true revivals" can be considered to be the ones, which nearly return to the original initial entanglement, $\varepsilon$, at certain instances. In this work, a narrow window of $\varepsilon-\varepsilon'$ is considered, such that $\varepsilon'$ is within $95\%$ of $\varepsilon$. Once the number of such true revivals, $\mathcal{F}$, are computed over an sufficiently long enough observation time, $t_{max}$, we divide $\mathcal{F}$ by $t_{max}$ and obtain the following measure as en effective quantification of environmental interruption
\begin{eqnarray}
\mathcal{N}= \mathcal{F}/t_{max}.
\label{non_marko_measure}
\end{eqnarray}
We would like to mention here that unlike most of the
measures of non-Markovianity discussed in literature, which computes the `amount' (or
`degree') of environmental perturbation by integrating the growth of entanglement during the course of evolution (see \cite{non_Marko_ent}), here, by counting the instances of {\it near-perfect} revivals, we are
computing the `frequency' (or `occasions') of its occurrence.  One of the motivations for counting such instances is to observe the effects of the distinct dynamical processes  in different regimes of the considered physical model on the revival of quantum properties of the qubits \cite{MBL_ETH_entanglement1}. In the case of the ergodic phase, the evolution of the subsystems can be interpreted as a resulting  combined effect of both dissipation and dephasing mechanism leading the subsystem towards thermal equilibration and wiping out the initial state memory in due process. Whereas, in the many-body localized phase, though the interaction-induced dephasing is present, the absence of any dissipation leads the subsystems to relax in a non-thermal state which can retain a memory of the initial conditions. However, the dephasing mechanism, which has been extensively studied to understand the unusual logarithmic growth of entanglement in many-body localized phase,  is expected to play an adverse role towards attaining near-perfect revivals of quantum properties. The absence of dissipation or dephasing in the Anderson localized phase prevents the subsystems from any kind of relaxation and it is expected that there would be frequent occasions when quantum properties would not only revive but also attain values very close to that of the initial state. In this way, the measure can play an instrumental role to distinguish different revival patterns in the localized phases.\

We compute the value of $\mathcal{N}$ averaged over $10^2$ number of disorder realizations and for time interval $t\in[0, 100$] and obtain the behavior of $\mathcal{N}$ with respect to the change of disorder strength $h$ in Figure ~\ref{ETH_MBL_Measure}. This provides us a scope to compare the frequency of environmental disruption to the  strength of the disorder: Higher the value of the disorder strength, more frequent is the interruption caused by the environment.
 We noticed that the measure $\mathcal{N}$ eventually saturates at very high value of the disorder strength. In the inset of Figure \ref{ETH_MBL_Measure}, for a low value of the disorder strength $h=1.0$, we plot the scaling of $\mathcal{N}$ with the total system size $N=S+E$. A clear dependence of $\mathcal{N}$ on the system size asserts the fact that at low values of disorder strength, the measure eventually goes to zero when a sufficiently large environment is considered.  In the ergodic phase of the total $N$-qubit system, the entanglement of the bulk exhibits volume-law growth. Hence, the system qubit $(S)$, as one may expect, gets more and more entangled with the rest of the qubits, and as a result its quantum correlation with the probe-qubit diminishes rapidly.  On the other hand, the many-body localized phase is characterized by a slow ( logarithmic ) spreading of entanglement. Thus the system qubit often gets  loosely entangled with the rest of the environment qubits and revival of system-probe entanglement takes place.   

\begin{figure}[h!]
\begin{center}
\includegraphics[width=2.8 in,angle=00]{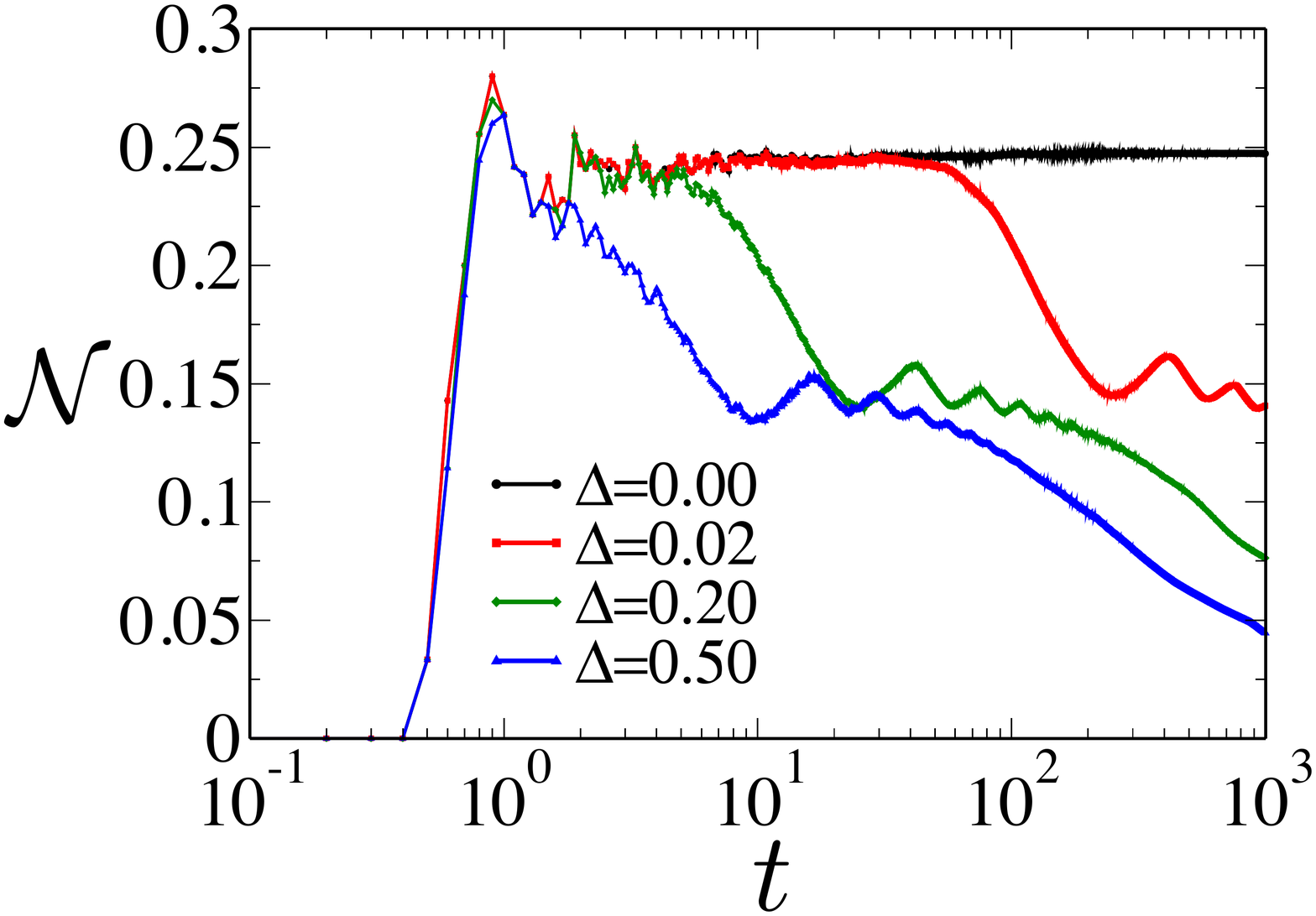}
\caption{(Color online.) Plot of the growth of the revival frequency ($\mathcal{N}$) in time ($t$)  obtained  for different $\Delta$ values viz., $\Delta=0.00$ (black), $\Delta=0.02$ (red),  $\Delta=0.20$ (green), and $\Delta=0.50$ (blue), for the disorder  strength $h=5$. The plots are obtained for  $N=10$ and for $10^2$ disorder realizations. }
\label{AL_MBL}
\end{center}
\end{figure}

Finally,  we aim to understand the role of many-body effects on  revival frequency $\mathcal{N}$. For this, we carry out similar analysis for the case when the interaction term in the Hamiltonian is absent i.e., $\Delta=0$. It is known from the literature that in this limit, the model reduces to a non-interacting model, which exhibits Anderson localization (AL) phenomena. Moreover, previous works in this direction report the emergence of quantum properties,  characteristically different  from the $\Delta\neq0$ case, which typically appear at  large time~\cite{MBL_ETH_entanglement1}. Motivated by these facts, we compute and compare the large time behaviors of the measure $\mathcal{N}$ in the interacting system and its non-interacting counterpart, and observe their growth in time. Figure \ref{AL_MBL} depicts the growth of $\mathcal{N}$ in time for different interaction strength, viz., $\Delta=0.00$ (black), $\Delta=0.02$ (red), $\Delta=0.20$ (green), and $\Delta=0.50$ (blue), for a moderately large disorder strength $h=5$. Interestingly, we note that the growth of the quantity $\mathcal{N}$ at the initial time (upto $t\approx 1$) remain independent of the interaction strength. The effects of interactions on revival frequency emerge at some later time. We observe that even for small non-zero interaction, $\mathcal{N}$ starts decaying beyond some critical time. The onset of such decay depends on the value of the interaction strength ($\Delta$) and can be approximated by $t_c\approx {1}/{\Delta}$. We note that the onset value reported here ($t_c$)  matches with the onset value obtained to mark the logarithmic growth of the entanglement entropy reported in earlier works~\cite{MBL_ETH_entanglement1}. Therefore, presence of interaction has an effect intermediate to the ergodic and Anderson localized phase: backflow of information occurs due to the absence of dissipation but dephasing hinders to attain perfect revivals of quantum properties. \\

 The behavior of revival frequency observed in these two localized phases has a close correspondence with that of the behavior of bulk ($S+E$) entanglement entropy obtained in this limit~\cite{MBL_ETH_entanglement1}. In the Anderson localized phase, entanglement entropy of the bulk saturates in time at a value much lower than that ergodic and the many-body localized phase. This results in weak entanglement of the system qubit with the rest of the $N-1$ qubits of the environment-part and strong revival of system-probe entanglement occurs with a rate that essentially saturates in time. However, in the presence of interaction, many-body effects come into the picture and at large times, the dephasing mechanism causes the logarithmic growth of the entanglement entropy of the bulk. The logarithmic decay of the system-probe
entanglement revival conveys an analogous onset of  
dephasing effect: backflow of information occurs due to the absence of dissipation but dephasing hinders to attain perfect revivals of quantum properties. Hence,  the rate of near-perfect revival becomes much lower than that of the Anderson localized phase. \\
 
  Our analysis provides a  fine-grained characterization of the localized phases of the model in terms of drastically different trends of information backflow in ergodic, Anderson and MBL phases. We would also like to mention here that similar analysis using well-known RHP-measure  \cite{non_Marko_ent}  of non-Markoniavity, a proposed measure for quantifying the total amount information backflow, fails to capture any such demarcations between these localized phases. In particular, the evolution pattern of the amount of backflow in Anderson and MBL phases remains practically indistinguishable even at large evolution times.\\

\section{Discussions}
\label{discussion}

In this work, using the tools of the theory of open quantum systems, we studied the trends of information backflow in the disordered transverse field Heisenberg model. From the perspective of the single qubit system-part, the evolution is realized via the action of a dynamical map acting on it. Subsequently,  we characterized the nature of the dynamical map by investigating the decay profiles of bipartite entanglement shared between the system and the probe-qubit. At very low values of the disorder strength, the monotonous decay profile of the initial entanglement reveals Markovian nature of the dynamics, which essentially asserts the absence of environmental interruptions. 
Eventually, at higher values of the disordered strength, we observe significant deviations of the decay profiles of the bipartite entanglement from that monotonic nature, which affirms frequent information backflow from the environment to the system or the non-Markovian nature of the dynamics. We then establish a relation between the environmental interruption to that of the disorder strength:   strongly many-body localized states of the environment interfere more in the entanglement dynamics of system and probe-qubit as compared to the extended states in weak disorder. 
Additionally, from the large time dynamics, we show that environmental interruption has a strong dependence on the interaction strength. For non-interacting case, after the initial growth, the revival frequency ultimately saturates in time. However,  in the presence of even small amount of interaction,  after the initial growth,  the revival frequency starts decaying beyond a critical value of evolution time.

\acknowledgements

The authors thank S. Bhattacharya for reading the manuscript and  A. Sen(De) and U. Sen for useful suggestions and discussions.  

\end{document}